\pdfoutput=1

\documentclass[11pt]{article}

\usepackage[preprint]{acl}

\usepackage{times}
\usepackage{latexsym}

\usepackage[T1]{fontenc}

\usepackage[utf8]{inputenc}

\usepackage{microtype}

\usepackage{inconsolata}

\newcommand\our{\makebox{\textsc{ProtLLM}}}

\newcommand\ourds{\makebox{InterPT}}
\usepackage{url}
\usepackage{enumitem}
\usepackage{xspace}
\usepackage{xcolor}
\usepackage{wrapfig}
\usepackage{caption}
\usepackage{amssymb}
\usepackage{multirow}
\usepackage{adjustbox}
\usepackage{thmtools}
\usepackage{thm-restate}
\usepackage{tabularx}
\usepackage{booktabs}
\usepackage{setspace}
\usepackage{colortbl}
\usepackage{amsmath}
\usepackage{graphicx}
\usepackage{bbding}

\usepackage{amsmath,amsfonts,bm}

\def\eqref#1{equation~\ref{#1}}

\def\1{\bm{1}}

\def\vtheta{{\bm{\theta}}}

\def\ve{{\bm{e}}}

\def\vh{{\bm{h}}}

\def\vv{{\bm{v}}}

\def\mW{{\bm{W}}}

\DeclareMathAlphabet{\mathsfit}{\encodingdefault}{\sfdefault}{m}{sl}
\SetMathAlphabet{\mathsfit}{bold}{\encodingdefault}{\sfdefault}{bx}{n}

\DeclareMathOperator*{\argmax}{arg\,max}

\definecolor{temp}{RGB}{0,0,255}   
\definecolor{todo}{RGB}{255,0,0}
\definecolor{highlight}{RGB}{218,165,32}
\definecolor{decay}{RGB}{192,192,192}
\definecolor{r1}{RGB}{25,95,225}
\definecolor{r2}{RGB}{30,144,255}
\definecolor{r3}{RGB}{135,206,235}
\definecolor{myblue}{RGB}{68, 114, 196}
\definecolor{myorange}{RGB}{237, 125, 49}

\title{\our{}: An Interleaved Protein-Language LLM with\\
Protein-as-Word Pre-Training}

\author{Le Zhuo$^{2}$\thanks{Equal contribution.} \quad Zewen Chi$^{1}$\footnotemark[1] \quad Minghao Xu$^{3}$\footnotemark[1] \quad Heyan Huang$^1$\thanks{Corresponding author.} \\ \textbf{Heqi Zheng$^4$ \quad Conghui He$^5$ \quad Xian-Ling Mao$^1$ \quad Wentao Zhang$^{3}$\footnotemark[2]} \\
$^1$School of Computer Science and Technology, Beijing Institute of Technology \\
$^2$Beihang University$\;\,$ 
$^3$Center for Machine Learning Research, Peking University$\;\,$ \\
$^4$State Grid Smart Grid Research Institute Co., Ltd. \\
$^5$Shanghai Artificial Intelligence Laboratory \\
\url{https://protllm.github.io/project}
}

\begin{document}
\maketitle

\begin{abstract}

We propose \textbf{\our}, a versatile cross-modal large language model (LLM) for both protein-centric and protein-language tasks. \our{} features a unique dynamic protein mounting mechanism, enabling it to handle complex inputs where the natural language text is interspersed with an arbitrary number of proteins. Besides, we propose the protein-as-word language modeling approach to train \our{}. By developing a specialized protein vocabulary, we equip the model with the capability to predict not just natural language but also proteins from a vast pool of candidates. Additionally, we construct a large-scale \textbf{inter}leaved \textbf{p}rotein-\textbf{t}ext dataset, named \textbf{\ourds}, for pre-training. This dataset comprehensively encompasses both (1) structured data sources like protein annotations and (2) unstructured data sources like biological research papers, thereby endowing {\our} with crucial knowledge for understanding proteins. We evaluate \our{} on classic supervised protein-centric tasks and explore its novel protein-language applications. Experimental results demonstrate that \our{} not only achieves superior performance against protein-specialized baselines on protein-centric tasks but also induces zero-shot and in-context learning capabilities on protein-language tasks.

\end{abstract}

\section{Introduction}

\begin{figure}[t]
\centering
\includegraphics[width=0.49\textwidth]{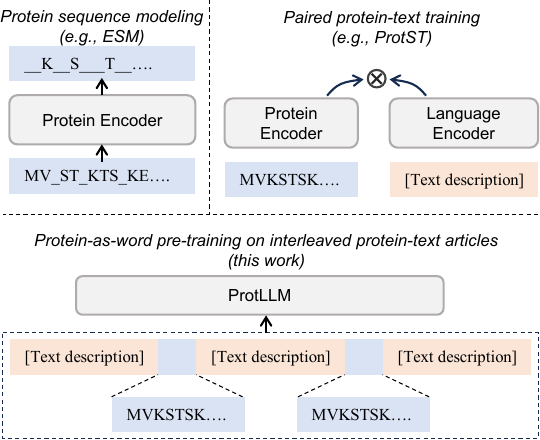}
\vspace{-1.5mm}
\caption{Unlike existing protein representation models that focus on protein-text pairs or protein-only data, \our{} can handle complex inputs with multiple proteins interleaved with text, thereby learning crucial knowledge from scientific papers and supporting diverse downstream tasks. }
 \label{fig:intro}
\vspace{-4.0mm}
\end{figure}

Understanding proteins is essential for unraveling the mysteries of life and enabling artificial intelligence systems to advance bioscience research \cite{wang2023scientific}. Thanks to the development of deep learning techniques, neural network models encompass extensive protein-centric applications, such as protein-folding prediction \cite{alphafold}, protein-protein interaction analysis \cite{li2018deep,su2023saprot}, function prediction \cite{zhang2023protein}, etc.

Protein representation learning methods typically employ large-scale pre-training, which learns unsupervised protein representations on massive protein sequences with masked language modeling \cite{rives2021biological}, or autoregressive language modeling \cite{elnaggar2020prottrans}.
In addition to protein-centric tasks, recent studies have attempted to extend protein models to protein-language scenarios.
ProtST~\cite{protst} integrates textual information into the protein encoder through multimodal pre-training on protein-text pairs, achieving zero-shot text-to-protein retrieval.
\citet{molinstruction} introduces an instruction dataset tailored for the biomolecular domain and investigates how fine-tuned LLM performs on protein-domain instruction-following tasks, such as function description generation.

Despite the success of protein representation methods on specific tasks, developing a model that excels in both protein-centric and protein-language tasks is still under-explored, facing three main challenges. Firstly, architectures are designed for particular downstream tasks, making it difficult to accommodate a wide range of tasks simultaneously. Secondly, current methods primarily derive cross-modal supervision from explicitly annotated protein-text pairs, which is not scalable to large-scale pre-training. Lastly, supporting a variable number of proteins in the input sequence introduces computational uncertainty in each training step, leading to inefficiencies during pre-training. 

In this work, we propose \our{}, which is a versatile LLM for both protein-centric and protein-language tasks. Instead of designing for specific tasks, \our{} supports complex interleaved protein-text inputs and outputs, which enables our model to simultaneously handle diverse downstream tasks without re-designing task-specific architecture (see Figure~\ref{fig:intro} for illustrations). Specifically, our dynamic protein mounting mechanism enables the model to seamlessly process text interspersed with an arbitrary number of proteins. Besides, we propose protein-as-word language modeling to ensure interleaved protein-text outputs. By building a protein vocabulary, \our{} is trained to autoregressively predict words and proteins from their respective vocabularies.

Additionally, we present a large-scale interleaved protein-text dataset, named \ourds{}, for \our{} pre-training. \ourds{} is constructed from diverse data sources, consisting of both structured data such as paired protein annotation data, and unstructured data from biological research papers, which encourages \our{} to harness crucial knowledge from the scientific articles.

We conduct extensive experiments on a wide range of downstream tasks, ranging from classic supervised protein-centric tasks to novel protein-language applications. Experimental results demonstrate that \our{} outperforms specialized baselines on protein-centric tasks.  \our{} also unlocks the in-context learning capability for protein-protein interaction prediction, and achieves zero-shot text-guided functional protein retrieval.

Our contributions are as follows:
\begin{itemize}
\item We propose \our{}, a versatile cross-modal LLM for both protein-centric and protein-language tasks. \our{} could process complex interleaved protein-text inputs and outputs, thereby supporting diverse tasks.
\item We introduce a large-scale pre-training dataset, \ourds{}, interleaving proteins and text from both structured data sources and unstructured multi-protein scientific articles. 
\item We show that \our{} achieves superior results on protein-centric tasks against protein-specialized baselines, and induces zero-shot and in-context learning capabilities.
\end{itemize}

\section{Related Work}
\subsection{Large Language Models}
The evolution of LLMs has been a cornerstone in the field of natural language processing~, showcasing extraordinary capabilities across a broad spectrum of tasks~\cite{bert,t5,gpt3,gpt4,llama,flan,palm}. These models, once thought to be limited to text-based tasks, have now crossed boundaries into areas traditionally dominated by human expertise, including mathematical problem-solving~\cite{cot,mathprompter}, drug discovery~\cite{drugchat,drugedit}, and complex decision making~\cite{yu2023language,eureka}. Recent explorations further extend LLMs' expertise into the multimodal domain where they demonstrate significant promise in processing and generating content from diverse modalities~\cite{Kosmos1,minigpt4,llava,graphtext,nextgpt}. Most of these works focus on aligning pre-trained encoders from various modalities with LLMs through instruction tuning, thus equipping LLMs to interpret multimodal inputs. In the realm of scientific research, specialized molecular LLMs have been devised for tasks like molecular property prediction~\cite{molxpt}, captioning~\cite{molinstruction}, and retrieval~\cite{molca}. Despite these advances, the progress in protein understanding with LLMs lags, hindered by the scarcity of comprehensive datasets for alignment and the absence of efficient architectures to model protein-language sequences.

\subsection{Protein Representation Learning}
Current mainstream methods for protein understanding tasks have focused on protein representation learning. Protein language models (PLMs)~\cite{elnaggar2020prottrans,rives2021biological,meier2021language,lin2022language} have marked significant progress in the area by training the protein sequence encoders on massive protein sequence data. 
Protein structure encoding methods aim to learn coarse-grained amino-acid-level representations~\cite{gligorijevic2021structure,fan2022continuous,zhang2023protein,xu2023eurnet} or fine-grained atom-level representations~\cite{hermosilla2020intrinsic,jing2021equivariant,zhang2023pre}. 
Despite the success in protein modeling, protein-related text data are left unexplored, which contains valuable supervision signals crucial for understanding proteins.
To enhance protein understanding with text supervision, OntoProtein~\cite{ontoprotein} leverages knowledge graphs, utilizing gene ontology annotations to implicitly enrich protein representation with textual information. ProtST~\cite{protst} integrates textual information into the protein encoder through multimodal pre-training on protein-text pairs, achieving zero-shot text-to-protein retrieval.

Mol-Instruction~\cite{molinstruction} introduces a comprehensive instruction dataset specialized for biomolecules and further fine-tunes LLMs on this dataset. Similarly, InstructProtein~\cite{instructprotein} improves the quality of instruction datasets by sampling protein-text pairs from a structured knowledge graph. This line of work focuses on aligning protein with human language using LLMs. However, a limitation of these approaches lies in their direct incorporation of protein sequences into LLMs as text, leading to suboptimal protein modeling due to the LLMs not being pre-trained on extensive protein sequence datasets. In contrast, \our{} provides a versatile framework that excels in both classic protein-centric tasks and novel protein-text applications.

\begin{figure*}[t]
\centering
\includegraphics[width=1.0\textwidth]{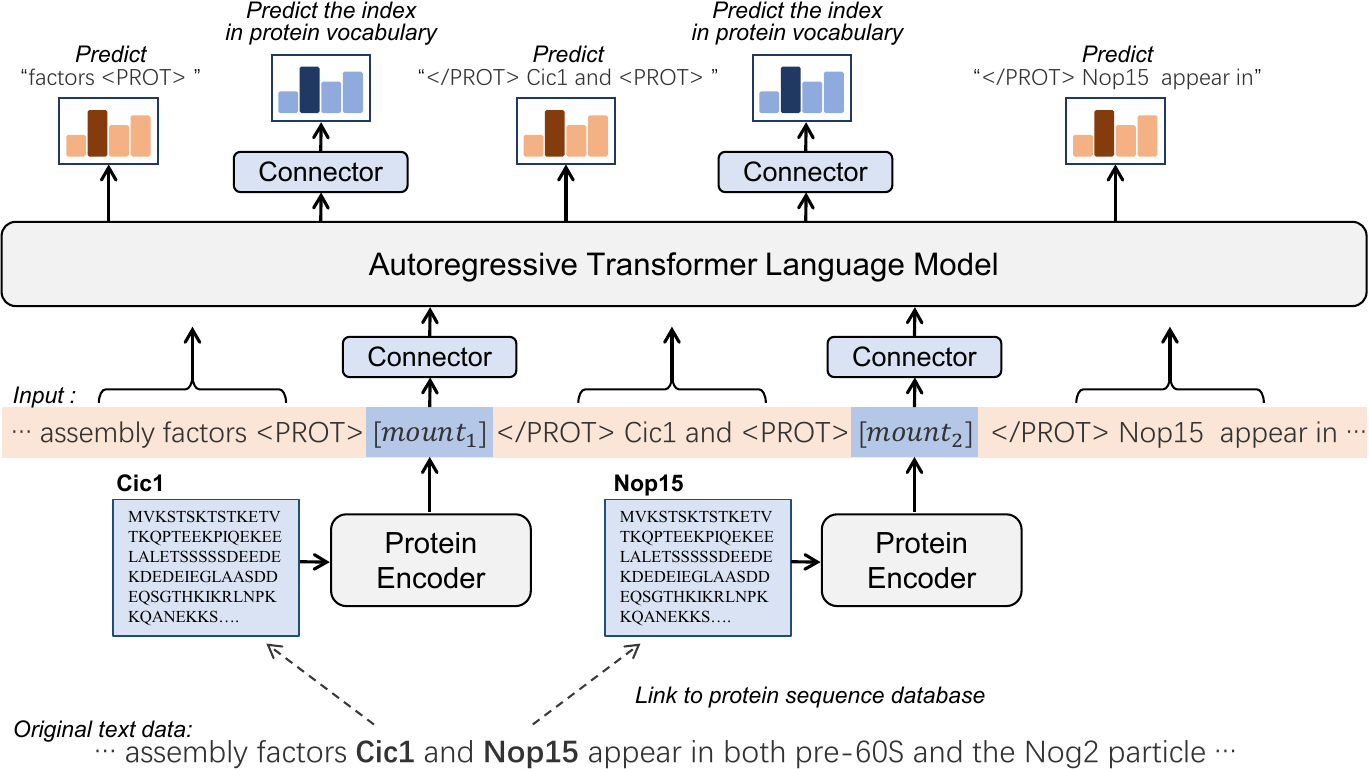}
\caption{An overview of \our{}. The architecture of \our{} consists of an autoregressive transformer, a protein encoder, and cross-modal connectors. With dynamic protein mounting, \our{} adeptly handles free-form interleaved protein-text sequences with an arbitrary number of proteins in the input. \our{} is pre-trained with protein-as-word language modeling that unifies word and protein prediction by constructing a protein vocabulary.}
 \label{fig:overview}
\end{figure*}

\section{Methods}

In this section, we elaborate on our proposed method, \our{}, which is illustrated in Figure~\ref{fig:overview}. Initially, we detail the model architecture in Section~\ref{sec:architecture}. Subsequently, the pre-training strategy is explained, introducing the concept of protein-as-word modeling, as outlined in Section~\ref{sec:pre-training}. We then present the uniquely constructed interleaved protein-text dataset, \ourds{}, in Section~\ref{sec:data}. Lastly, we explore the application of \our{} on a variety of tasks in Section~\ref{sec:application}.

\subsection{\our{} Framework}
\label{sec:model}

\paragraph{Model architecture}
\label{sec:architecture}
\our{} consists of an LLM for natural language modeling, a protein encoder, and cross-modal connectors that connect the protein encoder and the LLM.
We use LLaMA-7b~\cite{llama} as the backbone of \our{}, which is an autoregressive Transformer language model pre-trained on large-scale natural language data. To make \our{} understand protein sequences (i.e., sequences of amino acid tokens, which are the primary structure of proteins), we employ ProtST~\cite{protst} as the protein encoder. ProtST follows the backbone architecture of ESM-2~\cite{lin2022language} and introduces an additional two-layer MLP projection head. Pre-trained on large-scale protein-text pairs with contrastive learning, ProtST learns protein representations that are well-aligned with text. Besides, we introduce cross-modal connectors that connect the LLM with the protein encoder, thereby enabling \our{} to accept multimodal inputs. Specifically, \our{} has two cross-modal connector layers, which are placed at the input layer and the output layer of the LLM, respectively. The input-layer connector is a trainable projection matrix and transforms the output vectors from the protein representation space to the LLM representation space. Similarly, the output-layer connector transforms the LLM output vectors back to the protein representation space. Significantly, the output-layer connector also serves as a prediction head, allowing our model to perform protein retrieval and multi-choice protein answering tasks without requiring the LLM to generate complicated protein names.

\paragraph{Dynamic protein mounting}
\our{} considers not only structured protein-text paired data but also free-form interleaved protein-text sequences. Although the widely used encoder-decoder architecture can handle paired data, it encounters difficulties when dealing with interleaved protein-text inputs with multiple proteins. Therefore, we propose dynamic protein mounting, which allows \our{} to accept an arbitrary number of proteins as either input. Specifically, given an input sequence interleaved with proteins and text,
\begin{center}
\scalebox{0.75}{
    \texttt{...[text$_1$] [protein$_1$] [text$_2$] [protein$_2$] [text$_3$]...}
}
\end{center}
we do not directly feed the protein sequence to the LLM, but replace sequences with mount points.
\begin{center}
\scalebox{0.75}{
    \texttt{...[text$_1$] <PROT> [mount$_1$] </PROT> [text$_2$] ...}
}
\end{center}
At each mount point, we mount the protein encoder to the LLM with the cross-modal connector.  Additionally, these mount points are delineated by protein tags, signaling to the LLM that it is receiving protein vector inputs at these positions, rather than text data.

\subsection{\our{} Pre-Training}
\label{sec:pre-training}

\paragraph{Protein-as-word language modeling}
We introduce the protein-as-word language modeling training objective, which unifies protein prediction and word prediction as an autoregressive language modeling task. Consider an input sequence interleaved with $n$ tokens $[x_1, x_2, ..., x_n]$, where the $i$-th token $x_i$ represents either a natural language token or a protein. The protein-as-word language modeling object is to maximize the likelihood:
\begin{align}
    \argmax_\vtheta \sum_{i=1}^{n} \log p(x_i | x_{<i} ; \vtheta),
\end{align}
where $p(x_i | x_{<i}; \vtheta)$ is a categorical probability distribution over a natural language vocabulary when predicting natural words, or a protein vocabulary when predicting proteins. The probability is computed by
\begin{align}
\nonumber
&p(x_i | x_{<i} ; \vtheta) = \\ 
&\begin{cases}
\text{softmax}(\vh_i^\top\ve_j)_j, j \in \mathcal{V} ~~~~~~~~~~ \textit{if}~ x_i~\textit{is word} \\
\text{softmax}(\vh_i^\top\mW \vv_k)_k, k \in \mathcal{V}_\text{p} ~~~ \textit{if}~x_i~\textit{is protein}
\end{cases}
\end{align}
where $h_i$ is the last-layer LLM hidden states of $x_i$; $\ve_j$ is the word embedding of the word $j$ from the natural language vocabulary $\mathcal{V}$; $\mW$ stands for the output connector matrix, and $\vv_k$ is the protein embeddings of the protein $k$ from the protein vocabulary $\mathcal{V}_\text{p}$.
To construct the protein vocabulary, we collect all protein sequences in the training data. We then filter out proteins present in the downstream test sets to prevent data leakage. Finally, we compile a vocabulary consisting of the $1,076,781$ proteins.

\paragraph{Pre-training acceleration with protein cache}
Although our dynamic protein mounting design introduces flexibility for the input format, it also introduces computational uncertainty into the pre-training process, i.e., the computational cost of each step can vary significantly with the number of input proteins. Consequently, the throughput is limited by the worst case, leading to markedly reduced training efficiency. To accelerate the pre-training, we build a protein cache where we store all the pre-computed protein vectors encoded by the protein encoder. With the protein cache, we eliminate the heavy computational cost of the protein encoder, thereby accelerating the pre-training procedure with stable throughput. Besides, we utilize LoRA~\cite{hu2022lora} for efficient training.

\subsection{{\ourds}: Interleaving Protein-Text Data}
\label{sec:data}

We propose a large-scale \textbf{inter}leaved \textbf{p}rotein-\textbf{t}ext multimodal dataset, named \textbf{{\ourds}}, to pre-train {\our} with comprehensive protein-related knowledge. This dataset encompasses three types of data sources, i.e., multi-protein scientific articles, protein-annotation pairs, and protein instruction-following data. The statistics of each component are listed in Table~\ref{tab:stat}. 

\begin{table}[t]
\centering
\scalebox{0.83}{
\renewcommand\tabcolsep{4.5pt}
\begin{tabular}{llr}
\toprule
\textbf{Data Source} & \textbf{Data Type} & \textbf{Size} \\ \midrule
PubMed & Multi-protein articles & $165,206$  \\
UniProt & Annotations & $64,634$ \\
STRING & Annotations & $25,682$  \\
Mol-Instructions & Instruction-following data & $173,973$  \\
\bottomrule
\end{tabular}
}
\vspace{-1.0mm}
\caption{Category and statistics of {\ourds} components.}
\label{tab:stat}
\vspace{-2.0mm}
\end{table}

\paragraph{Multi-protein scientific articles} 
Multi-protein scientific articles describe complex relationships among different proteins found in biological research, where each sample could contain multiple proteins. Unlike data presented in structured formats such as pairs or knowledge graphs, these articles offer detailed insights in unstructured natural language. 
Guided by the recording in the STRING database~\cite{mering2003string} of multi-protein interactions and the scientific articles supporting them, we retrieve all involved articles from the PubMed database~\cite{canese2013pubmed}, specifically selecting instances where multiple proteins co-occur within the same paragraph. 
All proteins in these paragraphs are linked to the UniProt database~\cite{uniprot2015uniprot} for their amino acid sequences. Finally, we collect $165$K interleaved protein-text sequences from PubMed articles. 

\paragraph{Protein-annotation pairs}
This data maps individual proteins to their textual annotations such as function descriptions. We integrate two data sources, i.e., the UniProt database~\cite{uniprot2015uniprot} and the STRING database~\cite{mering2003string}, adding up to $90$K protein-annotation pairs. 
Given such a pair, we utilize it for two tasks, i.e., protein-to-text prediction and text-to-protein prediction, with the probability of $0.8$ and $0.2$, respectively. Besides, during pre-training, we interleave the data into longer sequences by concatenating multiple pairs into a single sequence, which has two advantages: (1) this operation can bridge the data length gap across different data sources and reduce the number of padding tokens, leading to higher training efficiency; (2) training multiple pairs in a single sequence encourages the model to obtain in-context learning capabilities~\cite{gu-etal-2023-pre}. 

\paragraph{Protein instruction-following data}
This data is in the instruction-following style~\cite{NEURIPS2022_b1efde53}, typically requiring the model to generate open-ended text given a protein and an instruction \cite{molinstruction}. We select the data items of proteins from the Mol-Instructions dataset~\cite{molinstruction} and include them into {\ourds}. Similar to the processing of protein-annotation pairs, we also concatenate multiple instruction-following data into a single pre-training example, so as to improve training efficiency and acquire in-context learning capabilities.

\subsection{Applying \our{} to Diverse Tasks}
\label{sec:application}

\paragraph{Supervised fine-tuning}
The best practice for adapting \our{} to downstream tasks is supervised fine-tuning when training data are available. Since \our{} supports flexible input and output formats, we can simply transform the downstream task data into an interleaved format and directly perform protein-as-word language modeling for supervised fine-tuning. The input and output prompt format for each downstream task can be found in the Appendix~\ref{app:detail}. During fine-tuning, we also apply the LoRA adapter to the LLM for efficient fine-tuning while preventing the model from overfitting several proteins in the training set.

\paragraph{In-context learning}
In-context learning is a promising capability of LLM, which can adapt the LLM to specific tasks with a few examples without training the model. \our{} can achieve in-context learning by pretending a few demonstration examples. To the best of our knowledge, \our{} is the first protein-language LLM that is capable of in-context learning.

\begin{table*}[t]
\begin{spacing}{1.1}
\centering
\begin{adjustbox}{max width=1\linewidth}
    \begin{tabular}{l|cc|cc|cc|cc|cc|c}
        \toprule
        \multirow{2}{*}{\bf{Model}} & \multicolumn{2}{c|}{\bf{Pre-training}} & \multicolumn{2}{c|}{\bf{EC}} & \multicolumn{2}{c|}{\bf{GO-BP}} & \multicolumn{2}{c|}{\bf{GO-MF}} & \multicolumn{2}{c|}{\bf{GO-CC}} & \bf{PPI} \\
        \cmidrule{2-3}
        \cmidrule{4-5}
        \cmidrule{6-7}
        \cmidrule{8-9}
        \cmidrule{10-11}
        \cmidrule{12-12}
        & Protein & Text & AUPR & $\mathrm{F}_{\mathrm{max}}$ & AUPR & $\mathrm{F}_{\mathrm{max}}$ & AUPR & $\mathrm{F}_{\mathrm{max}}$ & AUPR & $\mathrm{F}_{\mathrm{max}}$ & ACC \\
        \midrule
        DeepFRI & \Checkmark & \XSolidBrush & 0.546 & 0.631 & 0.282 & 0.399 & 0.462 & 0.465 & 0.363 & 0.460 & - \\
        GearNet & \Checkmark & \XSolidBrush & 0.892 & 0.874 & 0.292 & 0.490 & 0.596 & 0.650 & 0.226 & 0.486 & 73.86 \\
        ProtBert & \Checkmark & \XSolidBrush & 0.859 & 0.838 & 0.188 & 0.279 & 0.464 & 0.456 & 0.234 & 0.408 & 77.32 \\
        ESM-1b & \Checkmark & \XSolidBrush & 0.884 & 0.869 & 0.332 & 0.452 & 0.630 & 0.659 & 0.324 & 0.477 & 82.22 \\
        ESM-2 & \Checkmark & \XSolidBrush & 0.888 & 0.874 & 0.340 & 0.472 & 0.643 & 0.662 & 0.350 & 0.472 & 86.90 \\
        OntoProtein & \Checkmark & \Checkmark & 0.854 & 0.841 & 0.284 & 0.436 & 0.603 & 0.631 & 0.300 & 0.441 & 70.42 \\
        ProtST & \Checkmark & \Checkmark & \textbf{0.898} & \textbf{0.878} & 0.342 & 0.482 & 0.647 & \textbf{0.668} & 0.364 & 0.487 & 88.19 \\
        \bf{\our{}} & \Checkmark & \Checkmark & 0.874 & 0.860 & \textbf{0.349} & \textbf{0.503} & \textbf{0.652} & \textbf{0.668} & \textbf{0.469} & \textbf{0.596} & \textbf{89.87} \\
        \bottomrule
    \end{tabular}
\end{adjustbox}
\caption{Comparative benchmark results on protein-centric tasks. We use AUPR and $\mathrm{F}_{\mathrm{max}}$ on EC and GO prediction and accuracy (\%) on PPI prediction. Bold figures denote the best performance. `-' indicates not applicable.}
\label{tab:sft}
\end{spacing}
\end{table*}

\paragraph{Instruction-following protein retrieval}
For another interesting application, \our{} can be programmed to execute protein retrieval with customized requirements by following instructions. In Section~\ref{sec:case_study}, we show that \our{} can well retrieve functional proteins based only on function descriptions, and it can be further improved by prepending a one-shot demonstration.

\section{Experiments}
We evaluate \our{} on three types of downstream tasks: (1) protein-centric tasks, which include supervised fine-tuning on conventional benchmarks for protein understanding; (2) protein-text in-context learning, where we show the unique ability of \our{} by in-context learning on protein-protein interaction prediction; (3) text-guided functional protein retrieval, where we conduct a real-world enzyme mining task as a proof-of-concept study to validate the retrieval capability of \our{}. We present detailed hyperparameters, and prompt templates for pre-training and fine-tuning in Appendix~\ref{app:detail}.

\subsection{Protein-Centric Tasks}
\label{sec:sft}

\paragraph{Setup}
Following the settings in PEER benchmark~\cite{peer}, we adopt three standard tasks in protein understanding to validate our method. \textbf{Enzyme Commission (EC) number prediction}~\cite{gligorijevic2021structure} aims to predict all possible EC numbers of a protein simultaneously, reflecting the chemical reactions it catalyzes. \textbf{Gene Ontology (GO) term prediction}~\cite{gligorijevic2021structure} extends as a multi-label classification task, seeking to predict whether a protein belongs to specific GO terms. The GO benchmark is categorized into three branches, namely biological process (BP), molecular function (MF), and cellular component (CC). \textbf{Protein-Protein Interaction (PPI) prediction} aims to determine whether two given proteins interact or not. We adopt the human PPI dataset~\cite{pan2010large} for experiments.

To evaluate performances on multi-label classification tasks including EC and GO prediction, we report pair-centric area under precision-recall curve (AUPR) values and $\mathrm{F}_{\mathrm{max}}$, a widely used metric in the CAFA challenges~\cite{radivojac2013large}. PPI prediction results are evaluated by mean accuracy. These metrics require the soft probability of each target label. To achieve this, we initially extract the probabilities of ``Yes'' for the positive label and ``No'' for the negative label, respectively. Then, these probabilities are normalized via the softmax function to get the final predicted probabilities.

\paragraph{Baselines}
We compare ~\our{} with seven existing protein representation learning methods. As shown in Table~\ref{tab:sft}, these methods can be categorized into two distinct categories: \textbf{protein-only approaches} and \textbf{protein-text learning approaches}. The former encompasses sequence-based models including ProtBert~\cite{elnaggar2020prottrans}, ESM-1b~\cite{rives2021biological}, and ESM-2~\cite{lin2022language}, which are pre-trained using extensive collections of protein sequences, alongside structure-based models, such as DeepFRI~\cite{gligorijevic2021structure}, and GearNet~\cite{gearnet}. The latter, protein-text learning approaches, includes OntoProtein~\cite{ontoprotein}, and ProtST~\cite{protst}.

Note that Mol-Instructions~\cite{molinstruction} and InstructProtein~\cite{instructprotein} also belong to the protein-text learning approaches. However, their methods directly take protein sequences as human language and tokenize the data using byte-pair encoding. Contrasting with the protein-as-word strategy in \our{}, this exponentially increases the context length, rendering the evaluation on tasks with extensive label sets, like EC and GO prediction, or those requiring multiple protein inputs, such as PPI prediction, impractical for their approaches.

\paragraph{Results}
The results are shown in Table~\ref{tab:sft}. \our{} consistently shows competitive or even superior performance compared to both protein-only and protein-text approaches across all benchmarks, indicating the effectiveness of our proposed framework on conventional close-ended protein understanding tasks. Remarkably, \our{} obtain $0.596$ $\mathrm{F}_{\mathrm{max}}$  and $0.469$ AUPR on GO-CC, which outperforms ProtST by a large margin. As depicted in Section~\ref{sec:model}, \our{} directly uses pre-trained ProtST as the protein encoder, with the key difference lying in our LLM decoder and pre-training stage for alignment. By comparing \our{} with ProtST, the overall improvements strongly highlight the potential benefits of incorporating richer protein-text information and scaling the size of the language model.

Moreover, \our{} also outperforms two structure-based models on GO and PPI prediction despite we only leverage sequence information during training. Protein structures encode rich information and have direct relations to their functions. This opens up a promising direction to further incorporate protein structure into our framework, which we leave for future work.

\begin{figure}[t]
\centering
\includegraphics[width=0.42\textwidth]{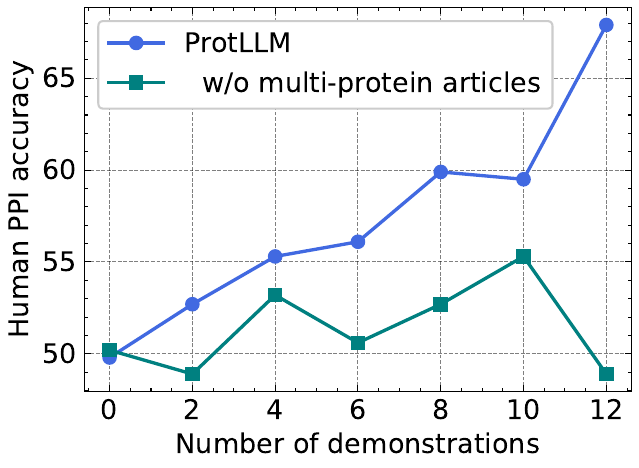}
\vspace{-1.5mm}
\caption{In-context learning results on human PPI.}
 \label{fig:icl}
\vspace{-1.0mm}
\end{figure}

\subsection{Unlocking In-Context Learning}

In-context learning is the capability that rapidly adapts the model to specific tasks using only a few annotated demonstration examples, which is originally found in autoregressive language models \cite{gpt3} and then is extended to visual language models \cite{flamingo}. In this section, we investigate whether \our{} can achieve in-context learning on the human protein-protein interaction (PPI) prediction task.

\paragraph{Setup} We directly evaluate the pre-trained \our{} model on the human PPI task without updating any parameters. For the $k$-shot in-context learning, we randomly sample $k$ examples from the validation set as the demonstrations and prepend them to each test sequence. Both the demonstration example and test example are prompted with the same template. For example, a one-shot prompted input is as follows:
\begin{center}
\scalebox{0.75}{
\begin{minipage}{0.63 \textwidth}
\texttt{Do <PROT> [mount$_1$] </PROT> and <PROT> [mount$_2$] </PROT> interact with each other? Yes\textbackslash n Do <PROT> [mount$_3$] </PROT> and <PROT> [mount$_4$] </PROT> interact with each other?}
\end{minipage}
}
\end{center}
\vspace{2mm}
The protein sequences of demonstration and test examples are first encoded by the protein encoder and then fed to the language model at each mount point. The final answer is predicted by selecting the verbalizer, i.e., ``Yes'' or ``No'', with the higher probability. Besides, to understand how multi-protein pre-training data from scientific articles improves \our{}, we also evaluate a variant of our model by removing the multi-protein scientific articles from the pre-training corpora.

\begin{figure*}[t]
\centering
    \includegraphics[width=1.0\linewidth]{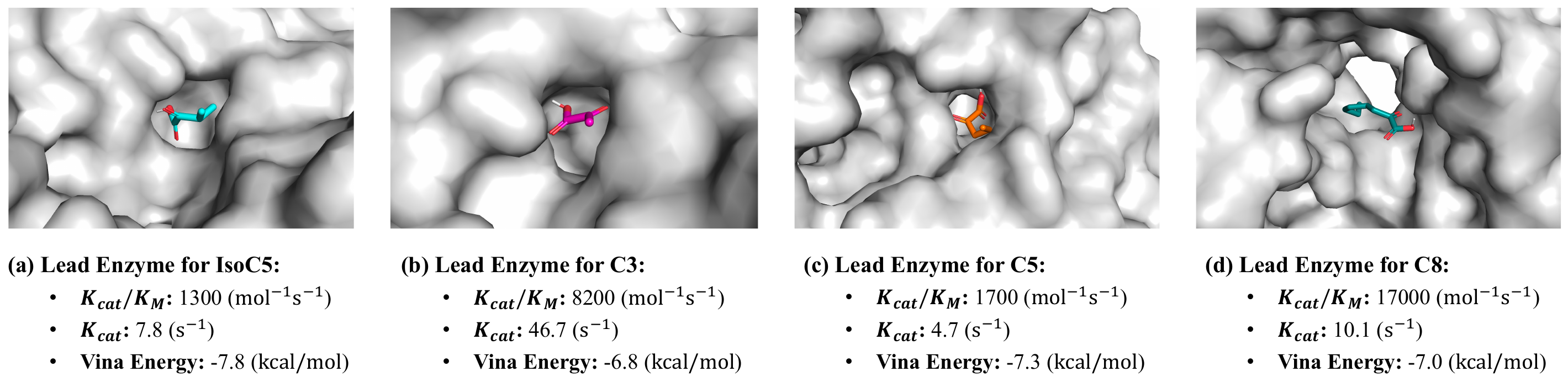}
    \caption{Top-1 enzyme mining results based on {\our} retrieval and AutoDock Vina post-screening. $K_{cat} / K_M$ and $K_{cat}$ measure enzyme activity (\emph{higher the better}). Vina energy measures binding affinity (\emph{lower the better}).}
    \label{fig:retrieval}
\vspace{-1.0mm}
\end{figure*}

\paragraph{Results}
Figure \ref{fig:icl} presents the in-context learning performance on human PPI with varying numbers of demonstration examples. Our model consistently achieves higher PPI accuracy with an increasing number of demonstration examples, demonstrating its effective in-context learning capability for protein-centric tasks. 
In comparison, the model performs drastically worse upon removing the multi-protein scientific articles, and fails to learn in context with the $2$, $6$, and $12$ demonstrations.
We believe that the in-context learning capability of our model could empower biologists to apply it to specialized tasks that lack annotated data, using minimal examples. Our experiments on enzyme mining illustrate a tangible application of in-context learning, as detailed in Section \ref{sec:case_study}.

\subsection{Text-Guided Functional Protein Retrieval}
\label{sec:case_study}

\begin{table}[t]
\begin{spacing}{1.1}
\centering
\vspace{-1.5mm}
\begin{adjustbox}{max width=1\linewidth}
    \begin{tabular}{l|l|ccc|ccc}
        \toprule
        \multirow{2}{*}{\bf{Reactant}} & \multirow{2}{*}{\bf{Method}} & \multicolumn{3}{c|}{\bf{Pool Size: 500}} & \multicolumn{3}{c}{\bf{Pool Size: 1000}} \\
        \cmidrule{3-8}
        &  & Top-10 & Top-20 & Top-50 & Top-10 & Top-20 & Top-50 \\
        \midrule
        \multirow{2}{*}{\bf{IsoC5}} & Zero-shot & 0.40 & 0.40 & 0.40 & 0.33 & 0.33 & 0.33 \\
        & In-context & \bf{0.60} & \bf{0.80} & \bf{0.80} & \bf{0.50} & \bf{0.67} & \bf{0.67} \\
        \midrule
        \multirow{2}{*}{\bf{C3}} & Zero-shot & \bf{1.0} & \bf{1.0} & \bf{1.0} & \bf{1.0} & \bf{1.0} & \bf{1.0} \\
        & In-context & \bf{1.0} & \bf{1.0} & \bf{1.0} & 0.67 & 0.67 & 0.67 \\
        \midrule
        \multirow{2}{*}{\bf{C5}} & Zero-shot & 0.40 & 0.40 & 0.40 & 0.25 & 0.25 & 0.25 \\
        & In-context & \bf{0.60} & \bf{0.80} & \bf{0.80} & \bf{0.38} & \bf{0.50} & \bf{0.50} \\
        \midrule
        \multirow{2}{*}{\bf{C8}} & Zero-shot & 0.33 & 0.33 & 0.50 & 0.22 & 0.22 & 0.33 \\
        & In-context & \bf{0.83} & \bf{0.83} & \bf{0.83} & \bf{0.56} & \bf{0.56} & \bf{0.56} \\
        \bottomrule
    \end{tabular}
\end{adjustbox}
\end{spacing}
\caption{Performance comparisons between zero-shot retrieval and in-context learning on enzyme mining. \textbf{Top-10, 20 and 50 Recall} are reported.}
\label{tab:retrieval}
\end{table}

\paragraph{Setup}
This experiment aims to study the capability of {\our} to retrieve functional proteins based on text prompts and demonstrations. For this purpose, we apply {\our} to enzyme mining, which is a critical stage in enzyme and metabolic engineering pipelines. In this experiment, we evaluate our model on mining carboxylate reductases that transform various ketoacids into their corresponding aldehydes. Four ketoacid reactants, i.e., 2-ketoisovaleric acid (IsoC5), pyruvic acid (C3), 2-ketovaleric acid (C5), and 2-ketooctanoic acid (C8), studied in \citet{mak2015integrative} are employed for evaluation. 

Using a reported enzyme for IsoC5, ketoisovalerate decarboxylase (KIVD)~\cite{de2004biochemical}, as the query, we first search for a pool of enzyme candidates by BLASTp~\cite{mcginnis2004blast}, where the pools with the size of 500 and 1000 are respectively tested. 
We then leverage {\our} to retrieve active enzymes from the pool for each reactant in two modes. In the \textbf{zero-shot retrieval} setting, given the prompt:
\vspace{-2mm}
\begin{center}
\scalebox{0.75}{
\begin{minipage}{0.65 \textwidth}
\texttt{Identify the enzymes: \{Reactant\} $\rightarrow$ Isobutanal. <PROT>}
\end{minipage}
}
\end{center}
\vspace{2mm}
describing the reaction from reactant (IsoC5, C3, C5 or C8) to product, {\our} generates a protein embedding at the token \texttt{<PROT>}. Then, we encode all the candidate enzymes as embeddings with the protein encoder. Finally, we utilize this embedding to rank all enzyme candidates by comparing embedding similarity. For \textbf{in-context learning}, we further add a one-shot demonstration of carboxylate reductase before the prompt above to facilitate enzyme mining. The demonstration is:
\vspace{-2mm}
\begin{center}
\scalebox{0.75}{
\begin{minipage}{0.65 \textwidth}
\texttt{Indentify the enzymes: Oxidizes aldehydes to the corresponding carboxylic acids with a preference for aromatic aldehydes. <PROT> [mount] </PROT>}
\end{minipage}
}
\end{center}
\vspace{2mm}
where the ``\texttt{[mount]}'' token is represented by the protein embedding of PaoC~\cite{neumann2009periplasmic}, a typical carboxylate reductase.

\paragraph{Results} 
In Table~\ref{tab:retrieval}, we report the recall of active enzymes found in \citet{mak2015integrative} at top 10, 20, and 50 ranked candidates. It is observed that in-context learning outperforms zero-shot retrieval on 18 out of 24 metrics, which verifies that {\our} can learn from a few demonstrations and improve its enzyme mining performance based on such knowledge. To study the top-ranked enzymes by {\our} more in-depth, we employ AutoDock Vina~\cite{trott2010autodock} to further screen the top-20 enzymes found by in-context learning and pick the one with the lowest Vina energy for visualization. As shown in Figure~\ref{fig:retrieval}, the lead enzymes selected in this way are all with good properties, possessing high enzyme activity (i.e., high $K_{cat} / K_M$ and $K_{cat}$ values measured by \citet{mak2015integrative}) and low binding energy measured by AutoDock Vina. These results altogether prove the effectiveness of {\our} on enzyme mining.

\section{Conclusion}
In this paper, we present \our{}, a versatile LLM designed to tackle both protein-centric and protein-language tasks. Through dynamic
protein mounting and protein-as-word modeling, \our{} adeptly handles complex interleaved protein-text data, seamlessly unifying a wide array of protein tasks via a natural language interface. Besides, we construct a large-scale protein-language pre-training dataset, called \ourds{}, which encourages the model to learn from diverse data sources ranging from structured paired data to unstructured multi-protein scientific articles. Extensive experiments demonstrate that \our{} not only achieves competitive performance against specialized baselines across standard protein-centric benchmarks but also paves the way for exploring novel protein-language applications.

\section*{Limitations}
In this paper, we primarily focus on sequence modeling for protein understanding. Nonetheless, \our{} is a general interface for the inputs in other modalities. Future research could further extend \our{} to additional modalities, such as protein structures and molecular graphs, by incorporating modality-specific encoders. Besides, we would like to explore more novel applications of \our{} such as scientific discovery.

\section*{Ethics Statement}
To the best of our knowledge, the \ourds{} dataset has been compiled from publicly available sources, carefully avoiding the inclusion of sensitive or private information. The primary focus of \our{} is to enhance protein understanding through various downstream tasks, distinguishing it from applications in dialogue systems. This focus inherently limits the potential for generating harmful content, leading to outputs that are inherently more controllable and safer. 
Nevertheless,  there remains a risk that malicious actors could exploit \our{} to spread misinformation or mislead users.

\section*{Acknowledgements}
We thank Jianan Zhao, Zuobai Zhang, Meng Qu, and Kunlun Zhu for the helpful discussions and comments.
\bibliography{acl_latex}

\begin{thebibliography}{55}
\expandafter\ifx\csname natexlab\endcsname\relax\def\natexlab#1{#1}\fi

\bibitem[{Alayrac et~al.(2022)Alayrac, Donahue, Luc, Miech, Barr, Hasson, Lenc, Mensch, Millican, Reynolds et~al.}]{flamingo}
Jean-Baptiste Alayrac, Jeff Donahue, Pauline Luc, Antoine Miech, Iain Barr, Yana Hasson, Karel Lenc, Arthur Mensch, Katherine Millican, Malcolm Reynolds, et~al. 2022.
\newblock Flamingo: a visual language model for few-shot learning.
\newblock \emph{Advances in Neural Information Processing Systems}, 35:23716--23736.

\bibitem[{Brown et~al.(2020)Brown, Mann, Ryder, Subbiah, Kaplan, Dhariwal, Neelakantan, Shyam, Sastry, Askell et~al.}]{gpt3}
Tom Brown, Benjamin Mann, Nick Ryder, Melanie Subbiah, Jared~D Kaplan, Prafulla Dhariwal, Arvind Neelakantan, Pranav Shyam, Girish Sastry, Amanda Askell, et~al. 2020.
\newblock Language models are few-shot learners.
\newblock \emph{Advances in Neural Information Processing Systems}, 33:1877--1901.

\bibitem[{Canese and Weis(2013)}]{canese2013pubmed}
Kathi Canese and Sarah Weis. 2013.
\newblock Pubmed: the bibliographic database.
\newblock \emph{The NCBI handbook}, 2(1).

\bibitem[{Chowdhery et~al.(2022)Chowdhery, Narang, Devlin, Bosma, Mishra, Roberts, Barham, Chung, Sutton, Gehrmann, Schuh, Shi, Tsvyashchenko, Maynez, Rao, Barnes, Tay, Shazeer, Prabhakaran, Reif, Du, Hutchinson, Pope, Bradbury, Austin, Isard, Gur-Ari, Yin, Duke, Levskaya, Ghemawat, Dev, Michalewski, Garcia, Misra, Robinson, Fedus, Zhou, Ippolito, Luan, Lim, Zoph, Spiridonov, Sepassi, Dohan, Agrawal, Omernick, Dai, Pillai, Pellat, Lewkowycz, Moreira, Child, Polozov, Lee, Zhou, Wang, Saeta, Diaz, Firat, Catasta, Wei, Meier-Hellstern, Eck, Dean, Petrov, and Fiedel}]{palm}
Aakanksha Chowdhery, Sharan Narang, Jacob Devlin, Maarten Bosma, Gaurav Mishra, Adam Roberts, Paul Barham, Hyung~Won Chung, Charles Sutton, Sebastian Gehrmann, Parker Schuh, Kensen Shi, Sasha Tsvyashchenko, Joshua Maynez, Abhishek Rao, Parker Barnes, Yi~Tay, Noam Shazeer, Vinodkumar Prabhakaran, Emily Reif, Nan Du, Ben Hutchinson, Reiner Pope, James Bradbury, Jacob Austin, Michael Isard, Guy Gur-Ari, Pengcheng Yin, Toju Duke, Anselm Levskaya, Sanjay Ghemawat, Sunipa Dev, Henryk Michalewski, Xavier Garcia, Vedant Misra, Kevin Robinson, Liam Fedus, Denny Zhou, Daphne Ippolito, David Luan, Hyeontaek Lim, Barret Zoph, Alexander Spiridonov, Ryan Sepassi, David Dohan, Shivani Agrawal, Mark Omernick, Andrew~M. Dai, Thanumalayan~Sankaranarayana Pillai, Marie Pellat, Aitor Lewkowycz, Erica Moreira, Rewon Child, Oleksandr Polozov, Katherine Lee, Zongwei Zhou, Xuezhi Wang, Brennan Saeta, Mark Diaz, Orhan Firat, Michele Catasta, Jason Wei, Kathy Meier-Hellstern, Douglas Eck, Jeff Dean, Slav Petrov, and Noah Fiedel. 2022.
\newblock {PaLM}: Scaling language modeling with pathways.
\newblock \emph{arXiv preprint arXiv:2204.02311}.

\bibitem[{Consortium(2015)}]{uniprot2015uniprot}
UniProt Consortium. 2015.
\newblock Uniprot: a hub for protein information.
\newblock \emph{Nucleic acids research}, 43(D1):D204--D212.

\bibitem[{De~La~Plaza et~al.(2004)De~La~Plaza, Fern{\'a}ndez~de Palencia, Pel{\'a}ez, and Requena}]{de2004biochemical}
Marta De~La~Plaza, Pilar Fern{\'a}ndez~de Palencia, Carmen Pel{\'a}ez, and Teresa Requena. 2004.
\newblock Biochemical and molecular characterization of $\alpha$-ketoisovalerate decarboxylase, an enzyme involved in the formation of aldehydes from amino acids by lactococcus lactis.
\newblock \emph{FEMS microbiology letters}, 238(2):367--374.

\bibitem[{Devlin et~al.(2018)Devlin, Chang, Lee, and Toutanova}]{bert}
Jacob Devlin, Ming-Wei Chang, Kenton Lee, and Kristina Toutanova. 2018.
\newblock {BERT}: Pre-training of deep bidirectional transformers for language understanding.
\newblock \emph{arXiv preprint arXiv:1810.04805}.

\bibitem[{Elnaggar et~al.(2020)Elnaggar, Heinzinger, Dallago, Rihawi, Wang, Jones, Gibbs, Feher, Angerer, Steinegger et~al.}]{elnaggar2020prottrans}
Ahmed Elnaggar, Michael Heinzinger, Christian Dallago, Ghalia Rihawi, Yu~Wang, Llion Jones, Tom Gibbs, Tamas Feher, Christoph Angerer, Martin Steinegger, et~al. 2020.
\newblock Prottrans: towards cracking the language of life's code through self-supervised deep learning and high performance computing.
\newblock \emph{arXiv preprint arXiv:2007.06225}.

\bibitem[{Fan et~al.(2022)Fan, Wang, Yang, and Kankanhalli}]{fan2022continuous}
Hehe Fan, Zhangyang Wang, Yi~Yang, and Mohan Kankanhalli. 2022.
\newblock Continuous-discrete convolution for geometry-sequence modeling in proteins.
\newblock In \emph{The Eleventh International Conference on Learning Representations}.

\bibitem[{Fang et~al.(2023)Fang, Liang, Zhang, Liu, Huang, Chen, Fan, and Chen}]{molinstruction}
Yin Fang, Xiaozhuan Liang, Ningyu Zhang, Kangwei Liu, Rui Huang, Zhuo Chen, Xiaohui Fan, and Huajun Chen. 2023.
\newblock Mol-instructions: A large-scale biomolecular instruction dataset for large language models.
\newblock \emph{arXiv preprint arXiv:2306.08018}.

\bibitem[{Gligorijevi{\'c} et~al.(2021)Gligorijevi{\'c}, Renfrew, Kosciolek, Leman, Berenberg, Vatanen, Chandler, Taylor, Fisk, Vlamakis et~al.}]{gligorijevic2021structure}
Vladimir Gligorijevi{\'c}, P~Douglas Renfrew, Tomasz Kosciolek, Julia~Koehler Leman, Daniel Berenberg, Tommi Vatanen, Chris Chandler, Bryn~C Taylor, Ian~M Fisk, Hera Vlamakis, et~al. 2021.
\newblock Structure-based protein function prediction using graph convolutional networks.
\newblock \emph{Nature communications}, 12(1):3168.

\bibitem[{Gu et~al.(2023)Gu, Dong, Wei, and Huang}]{gu-etal-2023-pre}
Yuxian Gu, Li~Dong, Furu Wei, and Minlie Huang. 2023.
\newblock \href {https://doi.org/10.18653/v1/2023.acl-long.267} {Pre-training to learn in context}.
\newblock In \emph{Proceedings of the 61st Annual Meeting of the Association for Computational Linguistics (Volume 1: Long Papers)}, pages 4849--4870, Toronto, Canada. Association for Computational Linguistics.

\bibitem[{Hermosilla et~al.(2021)Hermosilla, Sch{\"a}fer, Lang, Fackelmann, V{\'a}zquez, Kozl{\'\i}kov{\'a}, Krone, Ritschel, and Ropinski}]{hermosilla2020intrinsic}
Pedro Hermosilla, Marco Sch{\"a}fer, Mat{\v{e}}j Lang, Gloria Fackelmann, Pere~Pau V{\'a}zquez, Barbora Kozl{\'\i}kov{\'a}, Michael Krone, Tobias Ritschel, and Timo Ropinski. 2021.
\newblock Intrinsic-extrinsic convolution and pooling for learning on 3d protein structures.
\newblock \emph{International Conference on Learning Representations}.

\bibitem[{Hu et~al.(2022)Hu, yelong shen, Wallis, Allen-Zhu, Li, Wang, Wang, and Chen}]{hu2022lora}
Edward~J Hu, yelong shen, Phillip Wallis, Zeyuan Allen-Zhu, Yuanzhi Li, Shean Wang, Lu~Wang, and Weizhu Chen. 2022.
\newblock \href {https://openreview.net/forum?id=nZeVKeeFYf9} {Lo{RA}: Low-rank adaptation of large language models}.
\newblock In \emph{International Conference on Learning Representations}.

\bibitem[{Huang et~al.(2023)Huang, Dong, Wang, Hao, Singhal, Ma, Lv, Cui, Mohammed, Liu et~al.}]{Kosmos1}
Shaohan Huang, Li~Dong, Wenhui Wang, Yaru Hao, Saksham Singhal, Shuming Ma, Tengchao Lv, Lei Cui, Owais~Khan Mohammed, Qiang Liu, et~al. 2023.
\newblock Language is not all you need: Aligning perception with language models.
\newblock \emph{arXiv preprint arXiv:2302.14045}.

\bibitem[{Imani et~al.(2023)Imani, Du, and Shrivastava}]{mathprompter}
Shima Imani, Liang Du, and Harsh Shrivastava. 2023.
\newblock Mathprompter: Mathematical reasoning using large language models.
\newblock \emph{arXiv preprint arXiv:2303.05398}.

\bibitem[{Jing et~al.(2021)Jing, Eismann, Soni, and Dror}]{jing2021equivariant}
Bowen Jing, Stephan Eismann, Pratham~N. Soni, and Ron~O. Dror. 2021.
\newblock \href {https://openreview.net/forum?id=1YLJDvSx6J4} {Learning from protein structure with geometric vector perceptrons}.
\newblock In \emph{International Conference on Learning Representations}.

\bibitem[{Jumper et~al.(2021)Jumper, Evans, Pritzel, Green, Figurnov, Ronneberger, Tunyasuvunakool, Bates, {\v{Z}}{\'\i}dek, Potapenko et~al.}]{alphafold}
John Jumper, Richard Evans, Alexander Pritzel, Tim Green, Michael Figurnov, Olaf Ronneberger, Kathryn Tunyasuvunakool, Russ Bates, Augustin {\v{Z}}{\'\i}dek, Anna Potapenko, et~al. 2021.
\newblock Highly accurate protein structure prediction with alphafold.
\newblock \emph{Nature}, 596(7873):583--589.

\bibitem[{Li et~al.(2018)Li, Gong, Yu, and Zhou}]{li2018deep}
Hang Li, Xiu-Jun Gong, Hua Yu, and Chang Zhou. 2018.
\newblock Deep neural network based predictions of protein interactions using primary sequences.
\newblock \emph{Molecules}, 23(8):1923.

\bibitem[{Liang et~al.(2023)Liang, Zhang, Zhang, and Xie}]{drugchat}
Youwei Liang, Ruiyi Zhang, Li~Zhang, and Pengtao Xie. 2023.
\newblock Drugchat: towards enabling chatgpt-like capabilities on drug molecule graphs.
\newblock \emph{arXiv preprint arXiv:2309.03907}.

\bibitem[{Lin et~al.(2022)Lin, Akin, Rao, Hie, Zhu, Lu, dos Santos~Costa, Fazel-Zarandi, Sercu, Candido et~al.}]{lin2022language}
Zeming Lin, Halil Akin, Roshan Rao, Brian Hie, Zhongkai Zhu, Wenting Lu, Allan dos Santos~Costa, Maryam Fazel-Zarandi, Tom Sercu, Sal Candido, et~al. 2022.
\newblock Language models of protein sequences at the scale of evolution enable accurate structure prediction.
\newblock \emph{bioRxiv}.

\bibitem[{Liu et~al.(2023{\natexlab{a}})Liu, Li, Wu, and Lee}]{llava}
Haotian Liu, Chunyuan Li, Qingyang Wu, and Yong~Jae Lee. 2023{\natexlab{a}}.
\newblock Visual instruction tuning.
\newblock \emph{arXiv preprint arXiv:2304.08485}.

\bibitem[{Liu et~al.(2023{\natexlab{b}})Liu, Wang, Yang, Wang, Liu, Guo, and Xiao}]{drugedit}
Shengchao Liu, Jiongxiao Wang, Yijin Yang, Chengpeng Wang, Ling Liu, Hongyu Guo, and Chaowei Xiao. 2023{\natexlab{b}}.
\newblock Chatgpt-powered conversational drug editing using retrieval and domain feedback.
\newblock \emph{arXiv preprint arXiv:2305.18090}.

\bibitem[{Liu et~al.(2023{\natexlab{c}})Liu, Zhang, Xia, Wu, Xie, Qin, Zhang, and Liu}]{molxpt}
Zequn Liu, Wei Zhang, Yingce Xia, Lijun Wu, Shufang Xie, Tao Qin, Ming Zhang, and Tie-Yan Liu. 2023{\natexlab{c}}.
\newblock Molxpt: Wrapping molecules with text for generative pre-training.
\newblock \emph{arXiv preprint arXiv:2305.10688}.

\bibitem[{Liu et~al.(2023{\natexlab{d}})Liu, Li, Luo, Fei, Cao, Kawaguchi, Wang, and Chua}]{molca}
Zhiyuan Liu, Sihang Li, Yanchen Luo, Hao Fei, Yixin Cao, Kenji Kawaguchi, Xiang Wang, and Tat-Seng Chua. 2023{\natexlab{d}}.
\newblock Molca: Molecular graph-language modeling with cross-modal projector and uni-modal adapter.
\newblock \emph{arXiv preprint arXiv:2310.12798}.

\bibitem[{Longpre et~al.(2023)Longpre, Hou, Vu, Webson, Chung, Tay, Zhou, Le, Zoph, Wei et~al.}]{flan}
Shayne Longpre, Le~Hou, Tu~Vu, Albert Webson, Hyung~Won Chung, Yi~Tay, Denny Zhou, Quoc~V Le, Barret Zoph, Jason Wei, et~al. 2023.
\newblock The flan collection: Designing data and methods for effective instruction tuning.
\newblock \emph{arXiv preprint arXiv:2301.13688}.

\bibitem[{Ma et~al.(2023)Ma, Liang, Wang, Huang, Bastani, Jayaraman, Zhu, Fan, and Anandkumar}]{eureka}
Yecheng~Jason Ma, William Liang, Guanzhi Wang, De-An Huang, Osbert Bastani, Dinesh Jayaraman, Yuke Zhu, Linxi Fan, and Anima Anandkumar. 2023.
\newblock Eureka: Human-level reward design via coding large language models.
\newblock \emph{arXiv preprint arXiv: Arxiv-2310.12931}.

\bibitem[{Mak et~al.(2015)Mak, Tran, Marcheschi, Bertolani, Thompson, Baker, Liao, and Siegel}]{mak2015integrative}
Wai~Shun Mak, Stephen Tran, Ryan Marcheschi, Steve Bertolani, James Thompson, David Baker, James~C Liao, and Justin~B Siegel. 2015.
\newblock Integrative genomic mining for enzyme function to enable engineering of a non-natural biosynthetic pathway.
\newblock \emph{Nature communications}, 6(1):10005.

\bibitem[{McGinnis and Madden(2004)}]{mcginnis2004blast}
Scott McGinnis and Thomas~L Madden. 2004.
\newblock Blast: at the core of a powerful and diverse set of sequence analysis tools.
\newblock \emph{Nucleic acids research}, 32(suppl\_2):W20--W25.

\bibitem[{Meier et~al.(2021)Meier, Rao, Verkuil, Liu, Sercu, and Rives}]{meier2021language}
Joshua Meier, Roshan Rao, Robert Verkuil, Jason Liu, Tom Sercu, and Alexander Rives. 2021.
\newblock Language models enable zero-shot prediction of the effects of mutations on protein function.
\newblock \emph{bioRxiv}.

\bibitem[{Mering et~al.(2003)Mering, Huynen, Jaeggi, Schmidt, Bork, and Snel}]{mering2003string}
Christian~von Mering, Martijn Huynen, Daniel Jaeggi, Steffen Schmidt, Peer Bork, and Berend Snel. 2003.
\newblock String: a database of predicted functional associations between proteins.
\newblock \emph{Nucleic acids research}, 31(1):258--261.

\bibitem[{Neumann et~al.(2009)Neumann, Mittelst{\"a}dt, Iobbi-Nivol, Saggu, Lendzian, Hildebrandt, and Leimk{\"u}hler}]{neumann2009periplasmic}
Meina Neumann, Gerd Mittelst{\"a}dt, Chantal Iobbi-Nivol, Miguel Saggu, Friedhelm Lendzian, Peter Hildebrandt, and Silke Leimk{\"u}hler. 2009.
\newblock A periplasmic aldehyde oxidoreductase represents the first molybdopterin cytosine dinucleotide cofactor containing molybdo-flavoenzyme from escherichia coli.
\newblock \emph{The FEBS journal}, 276(10):2762--2774.

\bibitem[{OpenAI(2023)}]{gpt4}
OpenAI. 2023.
\newblock {GPT-4} technical report.
\newblock \emph{arXiv preprint arXiv:2303.08774}.

\bibitem[{Ouyang et~al.(2022)Ouyang, Wu, Jiang, Almeida, Wainwright, Mishkin, Zhang, Agarwal, Slama, Ray, Schulman, Hilton, Kelton, Miller, Simens, Askell, Welinder, Christiano, Leike, and Lowe}]{NEURIPS2022_b1efde53}
Long Ouyang, Jeffrey Wu, Xu~Jiang, Diogo Almeida, Carroll Wainwright, Pamela Mishkin, Chong Zhang, Sandhini Agarwal, Katarina Slama, Alex Ray, John Schulman, Jacob Hilton, Fraser Kelton, Luke Miller, Maddie Simens, Amanda Askell, Peter Welinder, Paul~F Christiano, Jan Leike, and Ryan Lowe. 2022.
\newblock \href {https://proceedings.neurips.cc/paper_files/paper/2022/file/b1efde53be364a73914f58805a001731-Paper-Conference.pdf} {Training language models to follow instructions with human feedback}.
\newblock In \emph{Advances in Neural Information Processing Systems}, volume~35, pages 27730--27744. Curran Associates, Inc.

\bibitem[{Pan et~al.(2010)Pan, Zhang, and Shen}]{pan2010large}
Xiao-Yong Pan, Ya-Nan Zhang, and Hong-Bin Shen. 2010.
\newblock Large-scale prediction of human protein- protein interactions from amino acid sequence based on latent topic features.
\newblock \emph{Journal of proteome research}, 9(10):4992--5001.

\bibitem[{Radivojac et~al.(2013)Radivojac, Clark, Oron, Schnoes, Wittkop, Sokolov, Graim, Funk, Verspoor, Ben-Hur et~al.}]{radivojac2013large}
Predrag Radivojac, Wyatt~T Clark, Tal~Ronnen Oron, Alexandra~M Schnoes, Tobias Wittkop, Artem Sokolov, Kiley Graim, Christopher Funk, Karin Verspoor, Asa Ben-Hur, et~al. 2013.
\newblock A large-scale evaluation of computational protein function prediction.
\newblock \emph{Nature methods}, 10(3):221--227.

\bibitem[{Raffel et~al.(2020)Raffel, Shazeer, Roberts, Lee, Narang, Matena, Zhou, Li, and Liu}]{t5}
Colin Raffel, Noam Shazeer, Adam Roberts, Katherine Lee, Sharan Narang, Michael Matena, Yanqi Zhou, Wei Li, and Peter~J. Liu. 2020.
\newblock Exploring the limits of transfer learning with a unified text-to-text transformer.
\newblock \emph{JMLR}.

\bibitem[{Rives et~al.(2021)Rives, Meier, Sercu, Goyal, Lin, Liu, Guo, Ott, Zitnick, Ma et~al.}]{rives2021biological}
Alexander Rives, Joshua Meier, Tom Sercu, Siddharth Goyal, Zeming Lin, Jason Liu, Demi Guo, Myle Ott, C~Lawrence Zitnick, Jerry Ma, et~al. 2021.
\newblock Biological structure and function emerge from scaling unsupervised learning to 250 million protein sequences.
\newblock \emph{Proceedings of the National Academy of Sciences}, 118(15).

\bibitem[{Su et~al.(2023)Su, Han, Zhou, Shan, Zhou, and Yuan}]{su2023saprot}
Jin Su, Chenchen Han, Yuyang Zhou, Junjie Shan, Xibin Zhou, and Fajie Yuan. 2023.
\newblock Saprot: Protein language modeling with structure-aware vocabulary.
\newblock \emph{bioRxiv}, pages 2023--10.

\bibitem[{Touvron et~al.(2023)Touvron, Lavril, Izacard, Martinet, Lachaux, Lacroix, Rozi{\`e}re, Goyal, Hambro, Azhar et~al.}]{llama}
Hugo Touvron, Thibaut Lavril, Gautier Izacard, Xavier Martinet, Marie-Anne Lachaux, Timoth{\'e}e Lacroix, Baptiste Rozi{\`e}re, Naman Goyal, Eric Hambro, Faisal Azhar, et~al. 2023.
\newblock Llama: Open and efficient foundation language models.
\newblock \emph{arXiv preprint arXiv:2302.13971}.

\bibitem[{Trott and Olson(2010)}]{trott2010autodock}
Oleg Trott and Arthur~J Olson. 2010.
\newblock Autodock vina: improving the speed and accuracy of docking with a new scoring function, efficient optimization, and multithreading.
\newblock \emph{Journal of computational chemistry}, 31(2):455--461.

\bibitem[{Wang et~al.(2023{\natexlab{a}})Wang, Fu, Du, Gao, Huang, Liu, Chandak, Liu, Van~Katwyk, Deac et~al.}]{wang2023scientific}
Hanchen Wang, Tianfan Fu, Yuanqi Du, Wenhao Gao, Kexin Huang, Ziming Liu, Payal Chandak, Shengchao Liu, Peter Van~Katwyk, Andreea Deac, et~al. 2023{\natexlab{a}}.
\newblock Scientific discovery in the age of artificial intelligence.
\newblock \emph{Nature}, 620(7972):47--60.

\bibitem[{Wang et~al.(2023{\natexlab{b}})Wang, Zhang, Ding, Qin, Zhuang, Li, and Chen}]{instructprotein}
Zeyuan Wang, Qiang Zhang, Keyan Ding, Ming Qin, Xiang Zhuang, Xiaotong Li, and Huajun Chen. 2023{\natexlab{b}}.
\newblock Instructprotein: Aligning human and protein language via knowledge instruction.
\newblock \emph{arXiv preprint arXiv:2310.03269}.

\bibitem[{Wei et~al.(2022)Wei, Wang, Schuurmans, Bosma, Ichter, Xia, Chi, Le, and Zhou}]{cot}
Jason Wei, Xuezhi Wang, Dale Schuurmans, Maarten Bosma, Brian Ichter, Fei Xia, Ed~Chi, Quoc Le, and Denny Zhou. 2022.
\newblock Chain-of-thought prompting elicits reasoning in large language models.
\newblock \emph{NeurIPS}.

\bibitem[{Wu et~al.(2023)Wu, Fei, Qu, Ji, and Chua}]{nextgpt}
Shengqiong Wu, Hao Fei, Leigang Qu, Wei Ji, and Tat-Seng Chua. 2023.
\newblock Next-gpt: Any-to-any multimodal llm.
\newblock \emph{arXiv preprint arXiv:2309.05519}.

\bibitem[{Xu et~al.(2023{\natexlab{a}})Xu, Guo, Xu, Tang, Chen, and Tian}]{xu2023eurnet}
Minghao Xu, Yuanfan Guo, Yi~Xu, Jian Tang, Xinlei Chen, and Yuandong Tian. 2023{\natexlab{a}}.
\newblock \href {https://openreview.net/forum?id=7fJC3tA1Ny} {Eurnet: Efficient multi-range relational modeling of protein structure}.
\newblock In \emph{ICLR 2023 - Machine Learning for Drug Discovery workshop}.

\bibitem[{Xu et~al.(2023{\natexlab{b}})Xu, Yuan, Miret, and Tang}]{protst}
Minghao Xu, Xinyu Yuan, Santiago Miret, and Jian Tang. 2023{\natexlab{b}}.
\newblock Protst: Multi-modality learning of protein sequences and biomedical texts.

\bibitem[{Xu et~al.(2022)Xu, Zhang, Lu, Zhu, Zhang, Chang, Liu, and Tang}]{peer}
Minghao Xu, Zuobai Zhang, Jiarui Lu, Zhaocheng Zhu, Yangtian Zhang, Ma~Chang, Runcheng Liu, and Jian Tang. 2022.
\newblock Peer: a comprehensive and multi-task benchmark for protein sequence understanding.
\newblock \emph{Advances in Neural Information Processing Systems}, 35:35156--35173.

\bibitem[{Yu et~al.(2023)Yu, Gileadi, Fu, Kirmani, Lee, Arenas, Chiang, Erez, Hasenclever, Humplik et~al.}]{yu2023language}
Wenhao Yu, Nimrod Gileadi, Chuyuan Fu, Sean Kirmani, Kuang-Huei Lee, Montse~Gonzalez Arenas, Hao-Tien~Lewis Chiang, Tom Erez, Leonard Hasenclever, Jan Humplik, et~al. 2023.
\newblock Language to rewards for robotic skill synthesis.
\newblock \emph{arXiv preprint arXiv:2306.08647}.

\bibitem[{Zhang et~al.(2022{\natexlab{a}})Zhang, Bi, Liang, Cheng, Hong, Deng, Zhang, Lian, and Chen}]{ontoprotein}
Ningyu Zhang, Zhen Bi, Xiaozhuan Liang, Siyuan Cheng, Haosen Hong, Shumin Deng, Qiang Zhang, Jiazhang Lian, and Huajun Chen. 2022{\natexlab{a}}.
\newblock \href {https://openreview.net/forum?id=yfe1VMYAXa4} {Ontoprotein: Protein pretraining with gene ontology embedding}.
\newblock In \emph{International Conference on Learning Representations}.

\bibitem[{Zhang et~al.(2022{\natexlab{b}})Zhang, Xu, Jamasb, Chenthamarakshan, Lozano, Das, and Tang}]{gearnet}
Zuobai Zhang, Minghao Xu, Arian Jamasb, Vijil Chenthamarakshan, Aurelie Lozano, Payel Das, and Jian Tang. 2022{\natexlab{b}}.
\newblock Protein representation learning by geometric structure pretraining.
\newblock \emph{arXiv preprint arXiv:2203.06125}.

\bibitem[{Zhang et~al.(2023{\natexlab{a}})Zhang, Xu, Jamasb, Chenthamarakshan, Lozano, Das, and Tang}]{zhang2023protein}
Zuobai Zhang, Minghao Xu, Arian~Rokkum Jamasb, Vijil Chenthamarakshan, Aurelie Lozano, Payel Das, and Jian Tang. 2023{\natexlab{a}}.
\newblock \href {https://openreview.net/forum?id=to3qCB3tOh9} {Protein representation learning by geometric structure pretraining}.
\newblock In \emph{The Eleventh International Conference on Learning Representations}.

\bibitem[{Zhang et~al.(2023{\natexlab{b}})Zhang, Xu, Lozano, Chenthamarakshan, Das, and Tang}]{zhang2023pre}
Zuobai Zhang, Minghao Xu, Aurelie Lozano, Vijil Chenthamarakshan, Payel Das, and Jian Tang. 2023{\natexlab{b}}.
\newblock Pre-training protein encoder via siamese sequence-structure diffusion trajectory prediction.
\newblock In \emph{Annual Conference on Neural Information Processing Systems}.

\bibitem[{Zhao et~al.(2023)Zhao, Zhuo, Shen, Qu, Liu, Bronstein, Zhu, and Tang}]{graphtext}
Jianan Zhao, Le~Zhuo, Yikang Shen, Meng Qu, Kai Liu, Michael Bronstein, Zhaocheng Zhu, and Jian Tang. 2023.
\newblock Graphtext: Graph reasoning in text space.
\newblock \emph{arXiv preprint arXiv:2310.01089}.

\bibitem[{Zhu et~al.(2023)Zhu, Chen, Shen, Li, and Elhoseiny}]{minigpt4}
Deyao Zhu, Jun Chen, Xiaoqian Shen, Xiang Li, and Mohamed Elhoseiny. 2023.
\newblock Minigpt-4: Enhancing vision-language understanding with advanced large language models.
\newblock \emph{arXiv preprint arXiv:2304.10592}.

\end{thebibliography}

\clearpage
\newpage
\appendix

\section{Experimental Details}
\label{app:detail}

\begin{table}[h!]
\centering
\scalebox{0.95}{
\begin{tabular}{lr}
\toprule
\textbf{Hyperparameter} &  \\
\midrule
Batch size & $256$ \\
Sequence length & $512$ \\
Training steps & $10$K \\
Optimizer & AdamW \\
Adam $\beta$ & ($0.9$, $0.999$) \\
Adam $\epsilon$ & $1\times10^{-6}$\\ 
Learning rate & $2\times10^{-4}$ \\
Learning rate schedule & Cosine decay \\
Warmup ratio & $0.03$ \\
Weight decay & $0$ \\
LoRA $r$ & $32$ \\
LoRA $a$ & $64$ \\
LoRA dropout & $0.1$ \\
LoRA modules & All linear modules \\
\bottomrule
\end{tabular}
}
\caption{
Pre-training hyperparameters of \our{}.
}
\label{tab:pretrian-hyper}
\end{table}

\begin{table*}[t]
\centering
\scalebox{0.95}{
\begin{tabular}{lrrrrr}
\toprule
\bf Hyperparameter & \bf EC & \bf GO-BP & \bf GO-MF & \bf GO-CC & \bf PPI \\
\midrule
Batch size & $128$ & $128$ & $128$ & $128$ & $16$ \\
Training steps & $50$K & $50$K & $50$K & $10$K & $10$K \\
Optimizer & AdamW & AdamW & AdamW & AdamW & AdamW \\
Learning rate & $2\times10^{-4}$ & $2\times10^{-4}$ & $2\times10^{-4}$ & $2\times10^{-4}$ & $2\times10^{-4}$ \\
Learning rate schedule & Cosine decay & Cosine decay & Cosine decay & Cosine decay & Cosine decay \\
Warmup ratio & $0.03$ & $0.03$ & $0.03$ & $0.03$ & $0$ \\
Weight decay & $0$ & $0$ & $0$ & $0$ & $0$ \\
LoRA $r$ & $128$ & $128$ & $128$ & $128$ & $32$ \\
LoRA $a$ & $256$ & $256$ & $256$ & $256$ & $64$ \\
LoRA dropout & $0.1$ & $0.1$ & $0.1$ & $0.1$ & $0.1$ \\
LoRA modules & All linear & All linear & All linear & All linear & All linear \\
Update protein encoder & Yes & Yes & Yes & Yes & No \\
\bottomrule
\end{tabular}
}
\caption{
Fine-tuning hyperparameters of \our{} on various downstream tasks.
}
\label{tab:finetune-hyper}
\end{table*}

\paragraph{Pre-training}
We list the detailed pre-training hyperparameters of \our{} in Table~\ref{tab:pretrian-hyper}. \our{} is pre-trained on $4$ NVIDIA A100 GPUs for $10,000$ steps with batch size $256$ on the \ourds{} dataset described in Table~\ref{tab:stat}. We adopt LoRA for efficient training, applying LoRA to all linear models of LLaMA, including~\texttt{[down\_proj, up\_proj, q\_proj, v\_proj, k\_proj, o\_proj, gate\_proj]}.
Notice that only the LoRA weights and the cross-modal connector modules are updated during training.

\begin{table*}[t]
\centering
\scalebox{0.8}{
\begin{tabular}{lll}
\toprule
\textbf{Task} & \textbf{Prompt template} & \bf Verbalizer  \\
\midrule
GO & \texttt{<PROT> [mount] </PROT> Does the protein belong to [name], which is [description]?} & \texttt{Yes/No} \\
EC & \texttt{<PROT> [mount] </PROT> Does the protein catalyze [name], which is [description]?} & \texttt{Yes/No} \\
PPI & \texttt{Do <PROT> [mount$_1$] </PROT> and <PROT> [mount$_2$] </PROT> interact with each other?} & \texttt{Yes/No} \\
\bottomrule
\end{tabular}
}
\caption{
Prompt templates for each task.
}
\label{tab:prompts}
\end{table*}

\paragraph{Fine-tuning}
\our{} is further fine-tuned on various downstream tasks including EC number prediction, GO term prediction, and PPI prediction. Table~\ref{tab:finetune-hyper} presents the fine-tuning hyperparameters. We apply LoRA for efficient tuning of the language model weights. The weights of the protein encoder are frozen for the PPI task, and updated for the other tasks, with a learning rate of $2 \times 10^{-5}$.
The handcrafted prompt templates for each task are shown in Table~\ref{tab:prompts}. At each \texttt{[mount]} position, we encode the protein sequences with the protein encoder and feed the resulting protein embedding to the language model. For multilabel classification tasks, i.e., GO and EC, we convert the tasks to binary classification tasks for each label. We fill \texttt{[name]} with the label name, and fill \texttt{[description]} with text descriptions associated with this label. Besides, we utilize a resampling strategy during fine-tuning to ensure a uniform distribution of positive and negative labels.

\end{document}